\documentclass[manuscript]{acmart}

\AtBeginDocument{%
  \providecommand\BibTeX{{%
    \normalfont B\kern-0.5em{\scshape i\kern-0.25em b}\kern-0.8em\TeX}}}

\usepackage{wrapfig}
\usepackage{censor}
\usepackage{enumerate}

\copyrightyear{2024}
\acmYear{2024}
\setcopyright{rightsretained}
\acmConference[FAccT '24]{The 2024 ACM Conference on Fairness, Accountability, and Transparency}{June 3--6, 2024}{Rio de Janeiro, Brazil}
\acmBooktitle{The 2024 ACM Conference on Fairness, Accountability, and Transparency (FAccT '24), June 3--6, 2024, Rio de Janeiro, Brazil}\acmDOI{10.1145/3630106.3658998}
\acmISBN{979-8-4007-0450-5/24/06}

\begin{document}

\title{Null Compliance: NYC Local Law 144 and the Challenges of Algorithm Accountability}

\author{Lucas Wright}
\authornote{Please direct all correspondence to Lucas Wright at law323@cornell.edu}
\email{law323@cornell.edu}
\orcid{0000-0002-3392-1928}
\author{Roxana Mika Muenster}
\email{rm858@cornell.edu}
\orcid{0000-0002-5052-5660}
\affiliation{%
  \institution{Cornell University}
  \city{Ithaca}
  \state{NY}
  \country{USA}
}


\author{Briana Vecchione}
\email{briana@datasociety.net}
\orcid{0000-0002-0828-8665}
\affiliation{%
  \institution{Data \& Society Research Institute}
  \city{New York City}
  \state{NY}
  \country{USA}
}

\author{Tianyao Qu}
\email{tq34@cornell.edu}
\orcid{0000-0002-5468-7376}
\author{Pika (Senhuang) Cai}
\email{sc3322@cornell.edu}
\orcid{0009-0004-3622-8952}
\affiliation{%
  \institution{Cornell University}
  \city{Ithaca}
  \state{NY}
  \country{USA}
}


\author{Alan Smith}
\email{alan.smith@consumer.org}
\orcid{0009-0009-2002-8071}
\affiliation{%
  \institution{Consumer Reports}
  \city{Yonkers}
  \state{NY}
  \country{USA}
}

\author{COMM/INFO 2450 Student Investigators}
\authornote{See Appendix \ref{studentinvestigators} for full list of student investigators.}
\affiliation{%
  \institution{Cornell University}
  \city{Ithaca}
  \state{NY}
  \country{USA}
}

\author{Jacob Metcalf}
\email{jake.metcalf@datasociety.net}
\orcid{0000-0002-2803-6625}
\affiliation{%
  \institution{Data \& Society Research Institute}
  \city{New York City}
  \state{NY}
  \country{USA}
}

\author{J. Nathan Matias}
\email{nathan.matias@cornell.edu}
\orcid{0000-0001-8910-0208}
\affiliation{%
  \institution{Citizens and Technology Lab, Cornell University}
  \city{Ithaca}
  \state{NY}
  \country{USA}
}

\renewcommand{\shortauthors}{Wright and Muenster, et al.}

\begin{abstract}
In July 2023, New York City became the first jurisdiction globally to mandate bias audits for commercial algorithmic systems, specifically for automated employment decisions systems (AEDTs) used in hiring and promotion. Local Law 144 (LL 144) requires AEDTs to be independently audited annually for race and gender bias, and the audit report must be publicly posted. Additionally, employers are obligated to post a transparency notice with the job listing. In this study, 155 student investigators recorded 391 employers’ compliance with LL 144 and the user experience for prospective job applicants. Among these employers, 18 posted audit reports and 13 posted transparency notices. These rates could potentially be explained by a significant limitation in the accountability mechanisms enacted by LL 144. Since the law grants employers substantial discretion over whether their system is in scope of the law, a null result cannot be said to indicate non-compliance, a condition we call "null compliance." Employer discretion may also explain our finding that nearly all audits reported an impact factor over 0.8, a rule of thumb often used in employment discrimination cases. We also find that the benefit of LL 144 to ordinary job seekers is limited due to shortcomings in accessibility and usability. Our findings offer important lessons for policy-makers as they consider regulating algorithmic systems, particularly the degree of discretion to grant to regulated parties and the limitations of relying on transparency and end-user accountability.
\end{abstract}

\ccsdesc[500]{Social and professional topics~Governmental regulations}
\ccsdesc[500]{Computing methodologies~Artificial intelligence}

\keywords{algorithm audit, compliance, transparency, bias}

\maketitle

\section{INTRODUCTION}
On the cusp of potential major changes to the AI regulation landscape in many jurisdictions, New York City implemented the world's first law mandating the conducting and publishing of algorithmic bias audits for commercial products in July 2023. NYC Local Law 144 (LL 144) mandates that any NYC-based private employer or City agency that deploys certain automated employment decision tools (AEDTs) in the hiring or promotion process must conduct a disparate impact study\footnote{LL 144 specifically names these audits as `bias audits,' but because they only measure a specific type of bias (disparate impact) against a narrow set of protected demographic classes (race and gender) we choose to use the term `disparate impact audit/study'. Please see Appendix A and Section 2.4 for a fuller description of the meaning of `disparate impact', how that is measured, and its relation to both LL 144 and the so-called `four-fifth’s rule.'} for race and gender features, and make the \textbf{audit report} available to the general public via their website (Appendix \ref{terms}). The employer must also provide a \textbf{transparency notice} to any job seeker, informing them about the use of an AEDT (Appendix \ref{lawhistory}). 

This law includes elements of interest to any government seeking to regulate algorithmic systems. For example, the third-party audit requirement creates a market for auditors, following practices in accounting \citep{soll_reckoning_2014}, pollution \citep{shimshack_economics_2014}, and compliance monitoring \citep{kaplow_optimal_1994}. The requirement of notice follows customs in employment law \citep{leibbrandt_equal_2018}. Yet the outcomes of these policy components have yet to be established in real-world algorithm governance.

This paper presents early analysis of the publicly-available outcomes of this historically-important algorithmic governance regime, two years after its passage and five months after it came into effect. Collaborating with 155 student investigators acting as model job seekers, we report what can be learned about employer compliance with the law, report qualitative and quantitative findings on the job seeker experience, and analyze the contents of published audit reports. Given the reported widespread usage of AEDTs and large workforce in NYC, we found surprisingly low rates of affirmative compliance with the required public audit reporting (5\%) and transparency notices (3\%) in our sample. Yet the accountability structures of LL 144 create a conundrum for investigators: the high level of discretion granted to employers to decide if their systems are in scope means that any null result cannot be considered non-compliance. Research can measure a compliance rate, but not a non-compliance rate. To make sense of this situation, we introduce the term \textbf{null compliance} to describe a situation in which non-compliance cannot be ascertained.\footnote{`Null compliance' has been used in other fields, such as accounting, biomedical research, and instrument calibration to describe the extreme end of non-compliance, meaning zero compliance, where compliance with a standard can be measured on a scale. Our usage is distinct: in transparency-driven algorithm regulation where compliance is a binary state, null compliance describes not being able to ascertain the state.} Null compliance does not name a third state between compliance and non-compliance that the employer may hold; rather, it describes the state of public knowledge \textit{about} compliance within this algorithmic transparency regime. Although this may be the first named example of null compliance in algorithmic governance, we  anticipate seeing further examples in other algorithmic accountability regimes that are primarily driven by strategies of transparency and consumer choice.

This law has created market opportunities for algorithm auditors and increased publicly-available information about some AEDTs used by NYC-based employers. Yet we find that the discretion afforded to employers by the law hinders the full potential utility of mandated transparency. Null compliance makes the extent of AEDT use by employers unclear, blunts the effectiveness of transparency requirements, undermines the choice structures intended to protect job seekers, and obfuscates any evidence about changes or reductions in discrimination over time. Taken together, these problems in a promising and landmark law make it impossible for researchers to know if LL 144 is reducing employment discrimination, increasing it, or making it harder to reliably monitor overall.

\section{REGULATING AUTOMATED EMPLOYMENT DECISION TOOLS IN NYC WITH LOCAL LAW 144}

LL 144 grew out of longstanding but troubled efforts to introduce a degree of transparency to the use of algorithmic systems by the municipality of New York City \citep{kirchner_new_2017, kaye_new_2019, cahn_first_2019}. These efforts shifted toward regulating commercial systems with potential civil rights impacts when in early 2020 City Commissioner Laurie Cumbo sponsored a statute requiring employers to disclose and audit machine learning systems used to assist in hiring  \citep{new_york_city_council_local_2021}. The law was enacted in late 2021, and it assigned the Department of Consumer and Workers Protection (DCWP) responsibility for rule-making and enforcement. The DCWP went through two rounds of rule-making and public comment—including two delays in the implementation date—before issuing final rules in April of 2023, with implementation in July of 2023 \citep{new_york_city_department_of_consumer_and_worker_protections_automated_2023}. The law covers employers that are located in NYC, including remote jobs that primarily report to a NYC office.  

\subsection{Implementation Details of LL 144}

The final version of LL 144 imposes two related but distinct obligations upon employers\footnote{The law also applies to employment agencies, which procure and sometimes hire candidates for employers. We found no audits from these agencies.}, who are typically end-users of these systems.\footnote{If the AEDT is built in-house the employer may be both end-user and developer. We were unable to confirm any such case in our survey.} First, employers must hire an independent audit provider (described as an expert with no financial stake in the company or the outcome of the audit) to conduct an annual disparate impact audit of the system and post it publicly on their website. This \textbf{audit report} must focus on features of race/ethnicity, gender, and the intersection of both as defined by the U.S. Equal Employment Opportunity Commission (EEOC). Second, the employer must provide job seekers with a \textbf{transparency notice} that their application will be analyzed by an AEDT, and provide job seekers with an opt-out mechanism to request a human review process if one is available or otherwise required by law.\footnote{The law specifies that employers are \textit{not} obligated to provide an alternative/human-only evaluation method \textit{unless} alternative methods are required by law, i.e., for medical or disability accommodations. Nonetheless, the theory of change behind LL 144 emphasizes job seeker agency to not apply to jobs with discriminatory, or any, AEDTs, and thus a form of opt-out is always available.} The transparency notice must either accompany the job posting or be sent to applicants via correspondence. No law requires disclosure of bias among human reviewers, so job seekers do not have information about the alternative option. 

The law specifies civil penalties for violators, ranging from \$500 to \$1500 per violation per day.\footnote{The text of the law does not specify if the violation is per system, per applicant, or per job posting, making the cost of non-compliance unclear.} Enforcement relies on complaints; the law does not provide the DCWP with proactive investigatory or discovery powers, nor does it grant a private right of action to job seekers. To the best of the authors’ knowledge, no complaints have been submitted to the DCWP at the point of publication. Due to the regulatee discretion detailed below, job seekers are unlikely to know whether any AEDT they may encounter during the application and interview should properly be in scope, and researchers and regulators cannot determine if the absence of an audit indicates absence of an AEDT.  





\subsection{Sources of Discretion Offered to Employers}

The drafters of LL 144 and the DCWP made several choices about the accountability structures of the law that significantly shape the conditions of this study, particularly the discretion granted to the employers that use AEDTs. \textbf{First}, LL 144 directly \textbf{imposes obligations upon employers only}, although most employers that use AEDTs lease them from vendors and/or recruitment platforms that train and maintain the models \citep{new_york_city_department_of_consumer_and_worker_protections_automated_2023}. As a municipality, NYC has regulatory powers over employers in its jurisdiction but not developers and platforms that are domiciled elsewhere. Additionally, employment anti-discrimination law focuses on moments of decision-making or evaluation between employer and employee, for which software platforms are typically not directly responsible.\footnote{This introduces significant challenges for an algorithmic accountability regime, however, because typically neither the end-user nor their hired auditor has direct access to the backend functionality and training data of the developer/vendor that are needed for algorithmic audits. Another paper using data from this study examines this issue in depth (see REDACTED, forthcoming).} \textbf{Second}, the DCWP establishes \textbf{no central public repository} for audit reports, relying solely on employers to post their own. \textbf{Third}, the law \textbf{does not bar systems} that cause disparate impact \citep{new_york_city_department_of_consumer_and_worker_protections_automated_2023}—it only requires reporting the rates. This choice places LL 144 in a complicated relationship to federal anti-discrimination law that does set a floor via the so-called `four-fifths rule' (see Appendices 1 and 2, and section 2.4 for more detail).

Given these features, the near-total discretion given to employers to interpret the scope of LL 144 has important consequences for  the success of the law. It is common practice to grant significant discretion to the regulatee to self-determine if and how the scope of a rule applies to them based on principles promulgated by the regulatory agency \cite{schmidt2021regulatory}. While LL 144 is not outside of typical parameters of regulatee discretion, it does appear to create a tension between discretion and transparency that ultimately blunts the effectiveness of the law, a result we name '\textbf{null compliance}'.

The DCWP defines an AEDT as software that uses machine learning techniques to "substantially assist or replace discretionary decision-making” to mean \citep{new_york_city_department_of_consumer_and_worker_protections_automated_2023}:

\begin{enumerate}[i.]
    \item to rely solely on a simplified output (score, tag, classification, ranking, etc.), with no other factors considered; or
    \item to use a simplified output as one of a set of criteria where the simplified output is weighted more than any other criterion in the set; or
    \item to use a simplified output to overrule conclusions derived from other factors, including human decision-making
\end{enumerate}

This is an awkward mix of technical features and organizational features that differs from technical definitions of machine learning \citep{barocas2023fairness}, introducing interpretational complexity and granting employers wide latitude over whether their systems are in scope. Given criteria ii. and iii., two employers that use identical models could grant those models different degrees of influence over their respective hiring decisions—an entirely organizational matter—and thereby have different regulatory statuses. Likewise, a job seeker could encounter the same vendor's AEDT during applications to two different employers and be protected by an audit at one but not the other, entirely contingent on how the employer describes their internal processes. Given those conditions, it is reasonable to expect that many employers and their legal advisors would be either confused about, or seek ways to legally avoid, the scope of the law. 

While there is no reliable source about what percentage of employers utilize AEDTs of any type, what surveys do exist show that AI systems for recruitment, hiring, and HR are widespread and rapidly growing in reach \citep{indeed_indeed_2023,shrm_automation_2022}. Vendors offer a wide variety of AI services to hiring managers, from matching candidates with openings to providing summaries of resumes to scheduling interviews; LL 144 applies only to AEDTs that are used at any point in the process to facilitate decisions regarding advancement/screening out of candidates \citep{new_york_city_department_of_consumer_and_worker_protections_automated_2023}. While ethically-complex use cases grab headlines (e.g., personality/psychological profiles, cognitive tests, and computer vision tools) lower-stakes AEDTs appear to be more common. Although LL 144 is built with them in mind, it does not appear that any vendors currently offer tools that fully replace human discretion in the hiring process with complete automation; thus any attempt to track the effect of the law would need to focus on systems that “substantially assist” humans.   

\subsection{Local Law 144 and Anti-Discrimination Law}

Finally, the relationship between LL 144 and federal anti-discrimination law is critically important to the outcomes of this study and the conundrum of null compliance. LL 144 requires an audit, but is silent on the results of that audit. The so-called `four-fifths rule'  \citep{bobko_four-fifths_2004, watkins_four-fifths_2022,code_of_federal_regulations_29_2024} that emerged from US federal anti-discrimination regulations and jurisprudence looms large in employment discrimination and algorithmic fairness (see \textbf{Appendix} for key terms and historical details). The four-fifths rule establishes a baseline against which impermissible discriminatory outcomes (aka `disparate impact') can be identified by comparing relative rates of selection between demographic groups. If the less-selected group's selection rate falls below four-fifths (80\%) of the baseline then there is a presumption that disparate impact may be occurring and warrants further investigation, typically through a complaint to the EEOC. The four-fifths rule is not genuinely a rule or law; it is best described as a rule of thumb or convention that is ingrained in hiring practices, and employers and developers are highly incentivized to reduce potential scrutiny by targeting that threshold. Nonetheless, LL 144 does not require that systems meet any discrimination threshold, including the four-fifths rule. Nor does LL 144 provide any guidance for remediation of systems when audits disclose disparate impact. The DCWP's FAQ notes that other laws on employment discrimination still apply but does not cross-reference them in any way \citep{new_york_city_department_of_consumer_and_worker_protections_automated_2023}.\footnote{The first case of algorithmic hiring bias—for age discrimination—was settled with the EEOC in 2023, but it was for a system that automatically screened out all older applicants and not a typical disparate impact scenario \citep{equal_employment_opportunity_commission_itutorgroup_2023, gilbert_eeoc_2023}.}

In other words, the most relevant statistical criterion for judging discrimination in employment is consequential to the first algorithm auditing law \textit{primarily by its absence}. This situation is central to understanding the conundrum of null compliance: one regulator (the DCWP) has demanded transparency about an activity that another regulator (the EEOC) has jurisdiction over but would not be able to observe in the usual course of business. The EEOC would typically need to pursue a discrimination complaint to find this information. Nor has any federal body established guidance about safe harbors for audits conducted in compliance with a local or state law. Such safe harbors would protect employers from liability for voluntarily contributing to transparency regimes. Thus, employers complying with LL 144 by posting audit reports might also be opening themselves to other types of liability, creating conflicting regulatory pressures  \citep{metcalf_what_2023}.  

That context speaks to LL 144’s theory of change, a term used by policy scholars to explain the link between policy details and its desired ends\citep{cairney2016politics, cairney2019understanding}. LL 144 requires transparent auditing so job seekers are empowered to make informed choices and employers and vendors are incentivized to deploy fair(er) models, but it does not set a floor for permissible rates of discrimination. It also creates a market for algorithm auditing services, and a sandbox for experimenting with auditing techniques.  The outcomes of LL 144 need to be measured against that theory of change.

\section{The Meaning of Null Compliance: Not the Same as Non-Compliance}\label{nullcompliance}

Because LL 144 gives employers so much discretion, any attempt to study compliance with the law will face difficulties estimating compliance rates due to \textit{presumed under-compliance and voluntary over-compliance}. The field of design differentiates between designers and end-users of systems who are not involved in their making \citep{norman2013design}. The law places accountability only upon the end-user of AEDT systems (the employers) to procure and post an audit, grants them near-total discretion over whether their system is in scope, and offers them many chances to move out of scope \citep{new_york_city_department_of_consumer_and_worker_protections_automated_faq_2023}. The law also offers no formal mechanism for challenging these employer decisions. Consequently, when investigators, regulators, and job seekers cannot find an audit report or transparency notice, they cannot call it non-compliance. 


To explain this situation, we introduce the term: `\textbf{null compliance}'. Null compliance describes a state in which the absence of evidence of compliance cannot be ascertained as non-compliance because the investigator lacks the information to determine if the regulated party’s actions or products are in scope of the regulation. Null compliance is a state of public knowledge that emerges from a transparency regime, not a state between compliance and non-compliance that the regulatee holds. In this situation, researchers can disclose an affirmative compliance rate (i.e., a regulatee affirms that they are in scope and compliant), but they cannot disclose a non-compliance rate. We anticipate seeing further examples in other algorithmic accountability regimes that emphasize transparency and yet offer discretion to end-users. 

In this paper, we report on information about whether researchers have observed an audit report and/or transparency notice on an employer's public web page. We use the term `null compliance' to refer to employers where our study was unable to verify compliance and also unable to verify non-compliance. Null compliance should not be taken to indicate non-compliance in this paper—absent legal investigatory powers that we lack, \textit{we cannot say that any given case of null compliance is actually non-compliance}.
A finding of a null compliance in our report may occur due to any combination of the following conditions made possible by LL 144’s particular accountability arrangements:

\begin{enumerate}
    \item The employer does not deploy an AEDT.
    \item The employer deploys AEDTs, but is not aware of LL 144.
    \item The employer deploys AEDTs, but has self-determined that their system is outside of the scope of the law.
    \item The employer deploys AEDTs, has self-determined that the law applies to their system, and is still searching for or waiting on a third-party auditor, which is a new industry. 
    \item The employer deploys AEDTs, has conducted an audit, and has decided that they should not post the results publicly as a risk management strategy.
    \item The employer deploys AEDTs, has conducted an audit, and complies with the law by providing job seekers with the transparency notice over email or the postal service rather than a public website.
    \item The employer deploys an AEDT and has conducted an audit, but has posted either/both the audit report and/or transparency notice in hard-to-discover location(s). 
\end{enumerate}

\section{Prior Scholarship on Measuring Compliance and Studying Opt-Out Decisions}
In this section, we summarize prior scholarship on measuring regulatory compliance and informed choice.



\subsection{Measuring Compliance with Transparency Regulations}\label{measuringcompliance}
Researchers in other fields have also wrestled with the difficulty of working with evidence produced through monitoring and transparency laws.  Because these policies are designed to influence regulated behavior while aiding enforcement \citep{shimshack_economics_2014}, they create incentives that complicate the study of regulated behavior. These incentives pressure firms toward under-compliance and over-compliance alike. The resulting uncertainty leaves society unclear about the outcomes of regulation. Our definition of null compliance incorporates an understanding of these sources of uncertainty.

\subsubsection{Deliberate Ignorance}
 Especially when research might expose firms to legal risks, ``trapped administrators'' can conclude that their self-interest depends on preventing or evading research \citep{campbell_reforms_1969}. In environmental policy and public health, scientists have described monitoring as a "co-evolutionary race" between researchers who seek knowledge and firms that seek to evade that research \citep{dietz_struggle_2003}. The threat to research validity from deliberate ignorance is so common that it is now the subject of a social science subfield on "agnotology" \citep{proctor_agnotology_2008}.

For researchers studying LL 144, deliberate ignorance could lead to under-counts in base rates of AEDT usage as well as biased estimates of racial and gender disparities in AEDTs. In the case of LL 144, employers have strong financial incentives to avoid conducting and publishing a bias audit. While the cost of non-compliance with LL 144 is a theoretical maximum of \$547,500 per year, civil action by federal regulators or private cases brought by litigators can be much larger \citep{equal_employment_opportunity_commission_itutorgroup_2023,gilbert_eeoc_2023}. Since non-compliance with LL 144 prevents third parties from knowing details that could expose employers to liability, they have a strong incentive toward deliberate ignorance.

\subsubsection{Information Access Control}

When regulated entities do choose to know things, they can still manage regulatory risk by limiting access to that information. Economists have described how self-reporting policies incentivize firms to publish results only when the costs and uncertainty of self-reporting are lower than those of withholding information \citep{kaplow_optimal_1994}. Access control, which economists describe in terms of information asymmetries, can create challenges for researchers seeking to estimate rates of compliance \citep{shimshack_economics_2014}. Computer scientists have described a similar dynamic where firms are incentivized to share data with researchers if they believe the results of research will be favorable \citep{young_confronting_2022}. For LL 144, access control could lead to an under-count of AEDTs, with mostly favorable audit reports published. 
 

\subsubsection{Mis-measurement and Fabrication}

Mis-measurement by regulated entities is a further threat to the validity of research on regulatory compliance. When regulations impose penalties on the basis of measurement, firms may respond by reporting false or adjusted information. In the study of air quality, for example, researchers have found systematic evidence of fabricated and false data from regulated firms and third-party auditors \citep{marchi_assessing_2006,short_integrity_2016,duflo_truth-telling_2013}. Scientists have also found evidence of systematic bias in official air quality measurements in cases where local politicians have discretion in their placement \citep{desouza_distribution_2021}. In the case of this study, we are unable to determine whether mismeasurement is occurring.

\subsubsection{Creative Behavior at Decision Rules}

Researchers have long observed that measurements become less reliable the more they are used for decision-making, especially when observations are close to a decision rule \citep{campbell_assessing_1979,goodhart_problems_1984}. For example, education researchers wrestle with the reliability of measurements that are used for decisions such as school allocation, graduation, and college admissions \citep{murnane_methods_2010,richard_by_2009}. Parallel scholarship has studied the problem of publication bias in science that result from creative efforts by scientists to navigate p-value thresholds \citep{ioannidis_what_2019}.

LL 144 includes two decision-rules that invite creative behavior. The four-fifths rule (Appendix \ref{terms}) is often used as a rule of thumb in employment discrimination law. Since companies have discretion over whether their hiring software is governed by the law, some companies might use their discretion under LL 144 to avoid EEOC investigation.
\subsubsection{Over-Compliance}

Researchers seeking to accurately measure regulated behaviors also face the problem of over-compliance. Without over-compliance, researchers could claim that compliance rates are under-counts, but over-compliance adds uncertainty in the opposite direction.

When regulators create a market for compliance monitoring, multiple actors have incentives toward over-compliance by following policies that may not apply to them or by complying to a degree that is not required \citep{shimshack_enforcement_2008}. Risk-averse firms can over-comply out of an abundance of caution, even when a law does not apply to them or when the scope of a recent law remains uncertain \citep{calfee_effects_1984}. Firms sometimes comply with monitoring and reporting regimes to grow their reputation compared to competitors \citep{engel_overcompliance_2006}. Finally, third party monitors market their commercial services to firms where auditing services are not legally required. Any of these incentives could lead some firms to over-comply with LL 144. While over-compliance might generate public goods by giving job seekers more choices and increasing the transparency of hiring software, it also undermines the accuracy of research on the use of AEDT systems and compliance with the law.

\subsubsection{Obfuscation and Unusability}

Even when firms comply with expectations that they warn people about potential risks from their products \citep{dillard_product_1955}, they exercise considerable agency over the visibility and accessibility of disclosures \citep{madden_duty_1986,solove_introduction_2012}. Obfuscation affects research when disclosures are made so inaccessible that researchers cannot find them.

Since LL 144 gives job seekers the right to request a costly alternative to the AEDT, employers have strong financial incentives to make disclosures difficult to find. Employers have limited guidance and no case law to inform the user experience. Consequently, researchers seeking out these disclosures will likely under-count them.

\subsection{Notice and Consent in Other Areas of Regulation}

Promoters of disclosure and consent policies argue that they enable individuals to make informed decisions in cases of significant information asymmetry. These policies enable the state to regulate firms indirectly by balancing those asymmetries \citep{thaler_libertarian_2003}. For this to work, people need to be sufficiently informed and able to make choices.

Research in behavioral science and computing has found that, at best, disclosure and consent policies only help some people under some circumstances and that they can backfire \citep{purmehdi_effectiveness_2017,ikonen_consumer_2020}. In employment law, behavioral scientists have found that equal opportunity and diversity statements can reduce applications from minority job seekers \citep{leibbrandt_equal_2018}. Such statements can also increase discrimination when they convince minority job-applicants to disclose information that subsequently activates biased hiring decisions \cite{kang_whitened_2016}. The disclose-and-opt-out model of LL 144 is also similar to many data privacy regulations. In this model, individuals are expected to read detailed information about a company's practices and make an informed decision. Computer scientists have observed that this model requires hundreds of hours per year per person \citep{mcdonald_cost_2008}. Even with time, people still struggle to make choices in their own interests \citep{solove_introduction_2012, solove_myth_2021}.

\subsection{Characteristics of Successful Notice and Consent}

Notice and consent depends on informed decision-making, which relies on the accessibility, comprehensibility, and usability of information. As many researchers have observed, regulation can create strong incentives for firms to undermine each of these characteristics.

First, information needs to be accessible. If job seekers cannot find transparency notices or bias audit reports, then they effectively have no choice, even if one is available in some technical sense. In other areas of technology policy, firms have tended to make disclosures inaccessible, requiring a "scavenger hunt" just to exercise one's rights \citep{habib_its_2020}.

Second, disclosures need to be comprehensible enough for a job seeker to make an informed choice \citep{zong_individual_2020}. Ongoing literature in bioethics, psychology, and design is exploring exactly what it means to make an informed decision \citep{rothman_strangers_2017,wilbanks_electronic_2020, zong_bartleby_2022, zong_data_2023}. At minimum, an informed decision is one in which people are able to reason about the possible consequences based on reliable information about the possible choices \citep{wilbanks_electronic_2020}.

Finally, the success of notice and consent policies depends on usability. A growing toolbox of "dark patterns" offers people the illusion of choice while steering people toward decisions that benefit the designers to their own detriment \citep{mathur_what_2021, caragay_beyond_2023}. These dark patterns are especially common in areas where firms and the public have competing incentives, including price discrimination and regulatory compliance \citep{mathur_dark_2019}.

\section{Methods}
In this paper, we set out to summarize how employers are publishing information to job seekers, describe the user experience they are creating for job seekers, and analyze the contents of the audit reports. To do so, we organized 155 student investigators to record public information they found on the websites of 391 employers in a manner modeled after job-seeking. We then validated the information, invited employers to correct any errors, and analyzed the resulting corpus. We have published our archive
as an open resource.\footnote{\textbf{\url{https://citizensandtech.org/2024-algorithm-transparency-law}}}

To inform the design of the full research protocol, we collected examples of bias audit reports a week after the law went into effect on July 5, 2023. By one month after the law came into effect, we had found 19 audit reports, 14 from employers plus 5 from software vendors not covered by LL 144. 
This pilot dataset was largely identical to a crowd-sourced dataset later assembled by civil society actors \citep{gerchick_tracking_2023}.

Our employer sample included 568 employers that had hired graduates of the classes of 2021 and 2022, based on data from the university careers office.\footnote{This sample is limited to the employers that hire Cornell students, which tend to be credentialed/professional roles. AEDT usage and transparency may be different in other sectors of the job market.} Employers were included if they hired a communication or information science major or hired at least two students from any major. We also included employers identified in our pilot study and employers that had been ranked as the top 100 internship providers in the country.\footnote{The full list of these employers can be found at: https://www.nationalinternday.com/2023-top-internship-programs} We excluded businesses in states well beyond NYC or no presence in the US for a total of 511. Our final list of 391 employers was a random sample from this list. 

\subsection{Modeling job seekers on Employer Websites}
Computer scientists studying regulatory compliance and user experience have typically taken two kinds of approaches. In one approach, researchers use software to crawl websites \citep{mathur_dark_2019}. Other researchers take the approach we used by organizing investigators to undertake large number of searches that model user behavior \citep{habib_its_2020}.

We recruited 155 student investigators through an extra credit opportunity in a large lecture class for Communication and Information Science students at Cornell University in New York State, where many graduates are hired by employers in NYC \footnote{Participants in this exercise who consented to sharing their names are acknowledged in Appendix \ref{studentinvestigators}}. In an hour-long session, we trained them to search for audit reports and transparency notices as if they were a job seeker. They searched the company’s website and profiles on LinkedIn and Indeed for no more than 30 minutes per employer to model the amount of time a dedicated job seeker might reasonably invest into the process. In order to protect against adverse consequences for the students, they did not submit applications.\footnote{This method could not observe transparency notices that were only supplied after someone applies.}

For each employer, investigators recorded whether the employer offered jobs in New York City at the time of their search. They also recorded evidence of any bias audit report and transparency notice they found, including screenshots, links, document files, and a description of the section of the website where they found the material. To minimize false negatives, we encouraged investigators to record any and all material that might plausibly be related to LL 144.

Student investigators collected data in two stages from October 24 through November 9, 2023. In the first stage, each investigator was randomly assigned one employer to research, assigning each company to two investigators. In the second stage, we assigned each investigator three more employers to research, with less overlap.

\subsection{Data Verification and Aggregation}

We next produced a list of verified reports and transparency notices from student investigator findings. Because we instructed investigators to be as inclusive as possible to minimize false negatives, they sometimes recorded other types of disclosures or statements of rights as being related to AEDTs. We reviewed the evidence they submitted with their results, checked the website if necessary, and removed false positives.

We also identified cases where employers using the same AEDT vendor published duplicate audit reports. In our analysis of employers, we count each employer that provided an audit report. When analyzing the audit reports themselves, we included each unique audit report once. In cases where an audit report reported impact ratios for multiple performance goals, we counted impact ratios separately since they report unique results.

\subsection{Survey of Employers}

We also contacted employers with two rounds of emails from Consumer Reports to further verify whether our research team might have missed public information about LL 144. This yielded two additional audit reports, out of 26 responses (Appendix \ref{surveyresponse}). False negatives in this study are not necessarily or solely an indication of researcher error: they may also illustrate the significant challenge of discovering these regulatory artifacts for an actual job seeker.

\section{Findings}

Based on the data collected from employers, we offer findings on observed compliance, the user experience created by employers, and the AEDT impact ratios published in audit reports.

\subsection{Summarizing Observed Compliance with Local Law 144}

Among 267 employers posting open jobs in NYC, we found 14 audit reports and 12 transparency notices (Table 1). Our sample included 124 employers that did not have a job opening in NYC at the time of the search. Of those companies, we found 4 bias audits and 1 transparency notice; such an outcome was expected because the transparency notices typically appear in conjunction with an active listing. Notably, only 11 employers published both an audit report and a transparency notice that met our understanding of the law's requirements, suggesting incomplete compliance among employers that have determined their use of AEDTs falls within the scope of the law.

\begin{small}
\begin{table*}[t]
\centering
\begin{tabular}{|c|c|c|c|c|c|}
\hline
                               &  & Audit & Audit & Transparency & Transparency\\ 
                               & Total & Report (N) & Report (\%) & Notices (N) & Notices (\%) \\ \hline
Employers listing NYC Jobs     & 267                       & 14             & 5\%              & 12                       & 4\%                       \\
Employers not listing NYC Jobs & 124                       & 4              & 3\%              & 1                        & 1\%                       \\ \hline
All employers                  & 391                       & 18             & 5\%              & 13                       & 3\%  \\
\hline
\end{tabular}
\caption{\label{summarydata}Across 391 employers, researchers found 18 audit reports and 13 transparency notices}
\end{table*}
\end{small}

While this study can count instances of compliance with LL 144, the problem of null compliance prevents third parties such as ourselves or job seekers from verifying employers' compliance with the law in a manner that advances the laws' transparency goals. Null compliance describes the phenomenon of not being able to interpret the absence of an audit report and/or transparency notice in our findings, which could result from at least seven possible causes that each have different legal implications (see Section 4). Given this, the absence of an audit report or transparency notice is not an indicator of  non-compliance on the part of any individual employer. Because the law affords a level of discretion to employers that makes the denominator impossible to know, we likewise cannot make claims about the percentage of employers in compliance with the law. Furthermore, while we can describe the audit reports that have been made public, the denominator problem prevents us from making general claims about AEDT systems in New York City, including any trends in algorithmic bias that might result. This distortion from null compliance in public knowledge about AEDTs is a direct result of the design of the accountability mechanism in LL 144.

In follow-up communications, employers explained how they exercised the discretion afforded by the law. Of the 26 firms that responded to our survey, 23 reported that LL 144 did not apply to them and 3 reported that LL 114 applied to them in some other way. Of the 3 employers, 2 supplied a document that met our definition of a bias audit. 
One company reported that while they do use AEDT systems more widely, they have chosen not to use these systems for NYC-based jobs in order to comply with LL 114. 

\subsection{The User Experience of Compliance with LL 144}

We report on the user experience of compliance with LL 144 on the basis of quantitative and quantitative findings from student investigators and an analysis of the materials that employers published. We consider three overall questions: the user experience of LL 144 compliance, the readability of bias reports, and the user experience of opting out. 

\subsubsection{The Inaccessibility of Transparency Notices and Audit Reports}

The theory of change behind LL 144 assumes that employers will make information accessible to job seekers who can make informed decisions. In this study, trained student investigators assessed whether the a motivated job applicant could locate the information that LL 144 mandates. Overwhelmingly, students could not do so, describing the experience as challenging, time-consuming, and frustrating.

Student investigators repeatedly reported the difficulty of locating the transparency notice and bias report on employer websites. Often, they were unable to locate transparency notices or bias audit reports within the 30-minute allotted search time, despite clicking through 17 pages and 19 pages on average, respectively. Employers that did publish the information placed it in very different parts of their websites. Disclosures sometimes appeared in the footer of the website or in the careers FAQ; other times, it was accessible through a drop-down menu or downloadable as a PDF. Given the challenges faced by trained student investigators, we expect that job seekers will only rarely encounter audit reports or transparency notices even when available.

\subsubsection{Ambiguity Amidst Other Disclosures}
Information about LL 144 was often placed among a larger list of other legal disclaimers and notices in confusing ways. Consequently, investigators struggled to identify whether a given notice fulfilled the requirements of LL 144 or some other regulation. Notices sometimes included unnecessary information such as ambiguous labels or vague mission statements. One student investigator reported: “I spent a lot of time sifting through information on the website, but I could not find it anywhere, despite the company claiming they value diversity, equity, and inclusion and minimizing unconscious bias.”

For student investigators, this cacophony was amplified by many similar documents and disclosures, especially on large employers’ websites: Privacy notices, financial audit reports, diversity statements and reports, or financial compensation bias audit reports were all mixed in AEDT transparency notices or bias audit reports. These reports were difficult to distinguish from each other even by trained investigators. Of records that student investigators thought might be audit reports 31\% were audit reports and 20\% were transparency notices as defined by LL 144. We therefore do not believe most job seekers could distinguish these statements under normal circumstances.

\subsubsection{Ambiguity Over the Applicability of LL 144}

Some employers demonstrated compliance while also claiming that the law did not apply. One perplexing example presented an AEDT bias audit report but prefaced it with an extensive disclaimer that, in the employers’ opinion, their tool did not actually constitute an AEDT as defined by LL 144.



In several cases, what the student investigators identified as a transparency notice was actually the opposite. Instead of disclosing the use of AEDTs, these statements informed job seekers that the company did not use AEDTs when hiring or that they do not solely rely on them in hiring decisions (which is not enough to exclude them from the law's scope). 

\subsubsection{The Usability of Audit Reports and Transparency Notices}

Once job seekers find audit reports, they need sufficient, clear information to make informed decisions. Our study identified 18 employer-provided bias audit reports, with some providing the same report as their AEDT vendor, leaving us with 13 unique reports.\footnote{The final rule permits employers to pool data from multiple employers’ data that use the same vendors’ model, as long as the employer also contributes their data to the pooled dataset. Yet since audit reports cannot and should not be the same between employers, they are required to publish unique reports. 
} 

None of these reports included enough explanation to inform job seekers or guide a decision to opt-out of applying. Audit reports were typically a PDF document with information about the audit provider, the employer, and the AEDT vendor, along with data tables (see Appendix \ref{terms}). They usually included the scoring rate, selection rate, and impact ratio across sex and race/ethnicity, with minimal explanation. Only three transparency notices mentioned the types of data collected and used in automated decision-making. 



\begin{figure}{}{}
\includegraphics[width=0.98\linewidth]{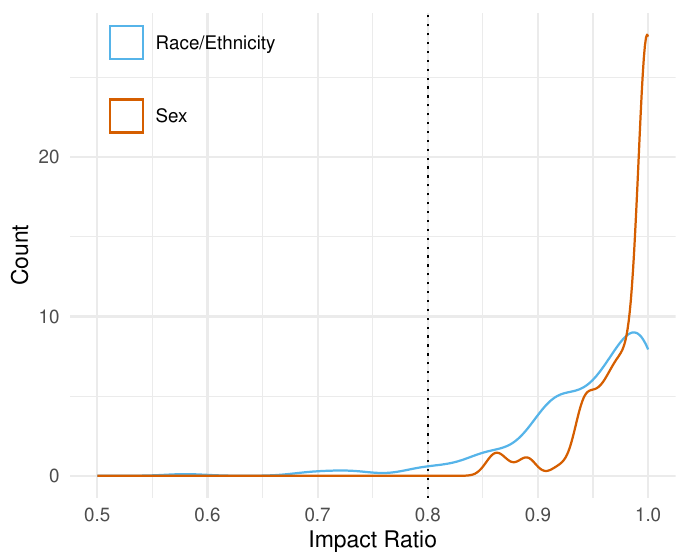}
\label{impactratios}
\caption{Kernel density showing the distribution of impact ratios for sex and race/ethnicity categories (N = 13 audit reports and 386 impact ratios) among publicly-posted audit reports}
\end{figure}

\subsection{Impact Ratios in Published Bias Audit Reports}


Most published audit reports reported impact ratios that were above an 0.8 threshold that would provide a presumption of non-discriminatory outcomes under federal law. While being on either side of the 0.8 threshold is not considered dispositive of either discrimination or non-discrimination in any jurisdiction, any numerical evidence of a disparate impact could draw negative attention and be used adversely against the employer in litigation (\ref{terms}, \ref{lawhistory}). As stated above, the `four-fifths rule' is correctly understood as a convention that is highly incentivized. Out of 386 total measurements, nine impact ratios were below 0.8, while all others reported impact ratios above this threshold.

We can only speculate on the reasons why 96\% of published audit reports included impact ratios above the 0.8 ``threshold”. It's possible that employers are more likely to share the results of favorable audits or that the estimates are mismeasured. It is also possible, though we think it is unlikely, that almost no employers in our sample used AEDT systems with performance below the 0.8 threshold. In a companion study to this one \cite{groves2024auditing}, audit industry workers indicated in interviews that, in their experience, many, if not the majority, of AEDTs on the market violate the four-fifths rule. Multiple audit service providers further indicated that they had clients who paid for an audit for an in-scope system but declined to post it publicly.

We did observe one employer that may be withholding an audit report. Since this employer unpublished their audit report before our study began, we did not included it. The report, which others have archived  \citep{gerchick_tracking_2023}, shows that the audit provider reported AEDT impact ratios below 0.8 in two demographic areas. We cannot explain why the report was unpublished—it is possible they ceased using AEDTs in NYC. When we contacted the employer, they did not respond. 

Why would any organization pay for an audit and then fail to comply with the law's public posting requirement? Because LL 144 does not set a floor for acceptable impact ratios,  employer compliance with LL 144 could make them vulnerable to enforcement from the EEOC or private litigation. Legal counsel may advise them that non-compliance with LL 144 is less risky than providing evidence for such litigation. While our data cannot be used to prove such publication bias in audit reports, our evidence is fully consistent with what we would expect in that situation and what auditors in Grove et al. \citep{groves2024auditing} indicated is the case.

The audit reports regularly excluded data for some race/ethnicity subgroups, offering the argument that the subgroup represented less than 2\% of the data being used for the audit, which is in line with the DCWP’s rules adopted to address concerns about distorted results due to low statistical power. Across all bias audit reports analyzed, employers omitted information for 20\% of impact ratios for this reason \citep{gaddis_audit_2018}. This is important to consider given that most omitted impact ratios fell under already marginalized indigenous groups. In one case, impact ratios were not reported at all: Holistic AI’s bias audit of Lendlease provided no results for impact ratios across race/ethnicity.

\section{Discussion}

The theory of change behind LL 144 is that transparency, the risk of EEOC enforcement, and job seeker agency will drive employers to adopt less biased systems and thereby create a market for well-governed and fair(er) AEDTs, or disincentivize use of such systems altogether. Policy-makers can learn important lessons about how LL 144's reliance on transparency and sustaining a market for auditing service does and does not make that possible. Ultimately, the effectiveness of the law should be judged on one outcome: reduced discriminatory outcomes in hiring and promotion where automated systems are deployed. Due to the problem of null compliance, the law and its current implementation do not enable anyone to know if that outcome is being achieved within individual employers or across the job market.

\subsection{Null Compliance in LL 144 Hinders Enforcement and Transparency}
Overall, the problem of null compliance hinders the law's function and prevents researchers from knowing if the law's purposes are being achieved. Because the law gives employers significant discretion (\ref{nullcompliance}), absence of information is null compliance rather than non-compliance, creating problems for enforcement and research alike. Null compliance creates strong barriers to enforcement. In such cases, investigators or complainant job seekers cannot tell whether the law does not apply, whether an employer is refusing to conduct a bias audit, or whether they are hiding an unfavorable audit. Null compliance also prevents researchers from compiling general information or claims about algorithmic hiring decision systems. The distribution of published impact ratios gives us reason to believe that publication bias could be causing a significant over-estimate of the fairness of AEDTs used to hire people in New York City.


Within audit reports, null compliance can obscure employment disparities for minority communities due to provisions in the law that permit employers to exclude data from audit reports. Since employers are in practice excused from reporting impact ratios for groups that they have hired the least, LL 144 is unable to achieve transparency or accountability for those groups. This sample size issue has important implications for algorithm accountability laws to address the rights of American Indians, Alaska Natives, and other indigenous groups.

The most significant change that could be made to LL 144 or subsequent laws would be to remove much of the discretion granted to the regulated party, which in this case is the end-user. In our judgment, the most effective route to this would be to attach the scope of the law to the purpose of the system. As it stands, employers can off-ramp from the regulatory decision tree by claiming (correctly or incorrectly) that their decision-making process does not `substantially' rely upon the outputs of the AEDT, or by using techniques that evade the technical definition of AEDTs. While this off-ramp may serve the purpose of incentivizing employers to reduce their reliance on AEDTs in decision making, it also creates the counter-intuitive outcome that a job seeker may encounter the same system across different employers but only be protected by LL 144 in some cases. Regulations could instead define AEDTs to include any system that outputs a simplified score or ranking from aggregated job seeker data, regardless of how the employer claims it is used or how it was built. This would reduce the regulatees' discretion (and confusion) over whether their AEDT is the right kind of AEDT to be regulated. If employers were required to publish consistently-formatted audit reports to a common repository, overall transparency would more easily be achieved.

\subsection{Confusing User Experiences and Partial Reporting Requirements Undermine job seeker Agency}

In theory, LL 144 empowers job seekers to make informed decisions about whether to accept an AEDT based on transparency notices and audit reports. Yet as the law was implemented by employers, job seekers cannot be reasonably expected to access, understand, or make use of that information. Transparency notices were hard to find and buried amidst many other disclosures. If job seekers do find the appropriate information, the legal and technical jargon in reports would prevent anyone except legal and technical experts from making an informed decision. Impact ratios and their meaning in anti-discrimination law are not simple to understand. Audit reports are not obligated to explain even basic concepts, such as that a lower score means a more discriminatory outcome.

A more effective regulation would provide clear, consistent guidance to employers on implementing notice and opt-out, including an opt-out for all job seekers. In other areas, federal regulators are considering `click to cancel' policies offering guidance on opt-out designs \citep{singletary_ftc_2023}. Even if employers published easy-to-understand audit reports with a simple user experience, job seekers still could not make truly informed decisions about AEDTs in the absence of information about the alternative. Looking at an impact ratio, the job seeker might rightly worry whether choosing human reviewers might expose them to even more bias than choosing AEDT. The most effective transparency regulation would require employers to report on the non-AEDT alternative as well. 
\subsection{Could LL 144 Reduce Employment Discrimination?}

In theory, publicly-available audit reports could provide evidence of changing impact ratios over time, guiding the evaluation of the overall outcomes of LL 144 on employment discrimination. Due to problems of null compliance and the lack of a mandated central repository for audit reports, tracking this is prohibitive. An improved law could set a floor for permissible impact ratios, even if that floor is fundamentally arbitrary and blunt. LL 144 does not set such a floor and instead relies on the `four-fifths rule' implicitly; it makes no reference to the rule but requires auditors to use exactly the equations and disclose results that any disparate impact investigation would make use of. This creates perverse incentives that make it less risky for companies to withhold data rather than prevent or reduce discrimination.

Absent some sort of safe harbor, any employer that uses a system with an impact ratio below 0.8 will need to decide if complying with LL 144’s transparency requirements will provide information for EEOC enforcement or private litigation. And because an impact ratio above 0.8 is not a guaranteed shield against litigation, publishing even apparently good results could conceivably feed adverse litigation outcomes. They may reasonably judge that it is highly unlikely that piecemeal enforcement by a local jurisdiction will be more costly than a federal civil suit.

\section{Conclusion}
Algorithmic employment tools are notoriously unreliable and ill-defined, imposing significant harm on workers \citep{narayanan_how_2019,raghavan_mitigating_2020,raji_fallacy_2022}. Policy experts have argued that the chief outcomes of disclosure-based regulations are the capacities they build to achieve at least some compliance \citep{selbst_institutional_2021} and facilitate grass-roots contestation \citep{metcalf_taking_2023}. Despite its significant flaws, Local Law 144 is achieving these first steps. In our judgment, the route by which LL 144 is most likely to drive change in the employment algorithm industry is by forcing the vendors’ customers to pause, measure and record outcomes, discuss internally about what their tools actually do, reflect on how they use them, and deliberate on whether their AEDTs are even necessary. We have some evidence that this is happening, from employers who claim to have stopped using these tools in New York City. If fewer employers use hiring algorithms, the rate of algorithmic bias will definitionally go down, even though it is hard to know if that would increase or decrease employment discrimination overall.

\begin{acks}
We are deeply grateful to the students of COMM/INFO 2450 for their thorough and extensive collaboration on this project and to the Cornell University Department of Communication for supporting this classroom research. J. Nathan Matias received financial support from the Center for Advanced Study in the Behavioral Sciences, the Lenore Annenberg and Wallis Annenberg foundation and the Siegel Family Endowment. Researchers at the Data \& Society Research Institute were supported by a grant from Omidyar Network, and in part by a grant from the Open Society Foundations. We thank our other collaborators on related projects, namely Alexandra Mateescu (who supported the pilot audit collection), Ranjit Singh, Brian Chen, Lara Groves, Andrew Strait, and Alayna Kennedy. We also thank the Data \& Society RAW Materials Seminar participants for their insightful comments. And we are grateful to the anonymous peer reviewers and conference committee members for supporting this venue for scholarship. 
\end{acks}

\bibliography{bibliography}
\bibliographystyle{ACM-Reference-Format}

\newpage

\section*{Ethics and Adverse Impact Statement}
\subsection*{Ethical Review}
We submitted this study for ethics review to Cornell University in protocol \#IRB0147978. Because the subjects in this study are organizations and the data is public, they concluded that the study does not qualify as human subjects research regulated under the Common Rule. We have nonetheless taken measures to minimize risk, contacting employers to give them a chance to correct any errors. We have also de-coupled the identities of student investigators from the dataset.

\subsection*{Adverse Impacts}
It is possible that this paper, when published, could influence and inform wider policy conversations alongside news coverage and enforcement of Local Law 144. Since this study focuses on organizations, we adopted best practices from journalism \citep{foreman_ethical_2022} to ensure the accuracy and reliability of our data, giving employers the opportunity to correct any errors.


\newpage 
\appendix
\section{Appendix}
\subsection{Key Terms for understanding employment discrimination}\label{terms}

\textbf{Selection Rate}: The rate at which members of a defined demographic group advance in a hiring process. ``Selection" is any decision point at which candidates are either chosen to move to the next stage—from resume screening to hiring—or rejected; it does not refer only to the final hiring decision.
\begin{center}
\vspace{8pt}
\textit{
    $\frac{\text{Number of applicants in a category}}{\text{Total number of applicants}}$} 
\vspace{8pt}
\end{center}

\textbf{Scoring Rate}: Scoring rate is the same concept as selection rate, but captures an important aspect of AEDTs generally: many AEDTs output a simplified \textit{scoring} or \textit{ranking} rather than a \textit{decision}. Examples include personality scores, culture fit scores, intelligence rankings, etc. The formula in LL 144 accounts for the rate at which the system scores applicants from the protected group with \textit{above median}, or desirable, scores.
\begin{center}
\vspace{8pt}
\textit{
    $\frac{\text{Number of applicants in a category with score > median score}}{\text{Total number of applicants in category}}$}
\vspace{8pt}
\end{center}

\textbf{Impact Ratio}: AKA impact factor, is a measure of the relative selections rates between one group and the most selected group. It measures differences in rates, not in absolute numbers. Identical selection rates result in an impact ratio of 1.0. The lower the number, the more discriminatory the selection process is. An impact ratio may also be calculated comparing the rates for the less-selected group against the rates for the entire population of candidates.
\begin{center}
\vspace{8pt}
    \textit{
    $\frac{\text{Selection rate for a category}}{\text{Selection rate of the most selected category}}$
    \hspace{12pt} OR \hspace{12pt} 
    $\frac{\text{Scoring rate for a category}}{\text{Scoring rate of the highest scoring category}}$}
\vspace{8pt}
\end{center}

\textbf{Disparate Impact}: AKA adverse impact, is an impact ratio that is low enough to be illegal or otherwise impermissible. In theory, disparate impact could be set at any possible impact ratio. In practice, it is largely defined by/identical to the four-fifths rule. Disparate impact is used to capture the phenomenon of unintentional discrimination through systemic features. An employer may be legally responsible for causing disparate impact even if none of their procedures explicitly discriminate against a protected group, e.g., if an employer has no policy against hiring a certain gender but their hiring practices result in \textit{relatively} lower rates of hiring for that gender, then they \textit{may} be guilty of causing disparate impact. Disparate impact may measure unintentional discrimination, but it does \textit{not} measure historical discrimination. The numerator may—and usually does in practice—correspond to a historically disadvantaged `protected category,' but it is not synonymous. Impact ratios can measure discriminatory outcomes against non-protected/historically-advantaged group. For example, an employer may have a policy of only hiring members of a historically disadvantaged group for a low-status job, resulting in a low disparate impact ratio when the historically disadvantaged group is the numerator (which would likely be illegal discrimination against both groups).  \\

\textbf{Four-Fifths Rule}: AKA 0.8 Rule or 80\% Rule, is a convention in US anti-discrimination law that defines disparate impact as any impact ratio that falls below 0.8, or four-fifths. In other words, if the rate at which a group is selected is lower than 80\% percent of the rate at which the most-selected group is selected, then the impact is presumed `disparate' or `adverse.' The `rule' in the name is a misnomer, it is instead a guideline that grants some degree of protection against regulatory scrutiny and litigation for hiring practices that result in impact ratios above 0.8. Falling below an impact ratio of 0.8 is not considered to be automatic evidence of illegal discrimination absent an investigation that considers many other factors that may legally justify a discriminatory outcome, such as legitimate relevance of the features to the job. However, falling below an impact ratio of 0.8 is a strong signal that a hiring process \textit{could} receive scrutiny and requires additional justification. The EEOC defines the Four-Fifths Rule in CFR §1607.4(D)\footnote {Available here: https://www.ecfr.gov/current/title-29/subtitle-B/chapter-XIV/part-1607/subject-group-ECFRdb347e844acdea6}.

\begin{center}
\vspace{8pt}
    \textit{
    $\frac{\text{Selection rate for a category}}{\text{Selection rate of the most selected category}} \leq 0.8$}
\vspace{8pt}
\end{center}

\subsection{Key Legal and Historical Details} \label{lawhistory}
Despite being a municipal law, the relationship between LL 144 and federal anti-discrimination law is critically important to the outcomes of this study. The so-called ``four-fifths rule”  \citep{bobko_four-fifths_2004, watkins_four-fifths_2022,code_of_federal_regulations_29_2024} that emerged from US anti-discrimination law looms large in employment jurisprudence and algorithmic fairness metrics. The ``four-fifths rule” is a general guideline that an impact ratio greater than 0.8 is considered \textit{de facto} evidence (but not \textit{de jure} evidence) of non-discriminatory outcomes. The concept of disparate impact and the four-fifths rule loom large in discussions of algorithmic bias because the unintentional, systemic discrimination described by disparate impact is often considered structurally analogous to algorithmic discrimination \citep{barocas_big_2016}. Indeed, in many algorithmic bias detection and mitigation tools, the four-fifths rule is treated as a foundational, standard measure of the line between acceptable and unacceptable discrimination, despite the fact that it is domain-specific inside of US law. As critics have noted \citep{watkins_four-fifths_2022}, the four-fifths rule is a fundamentally arbitrary threshold not supported by robust empirical evidence. But it is historically important, well-litigated, and embedded in U.S. anti-discrimination jurisprudence, and therefore familiar and unavoidable. 

LL 144 makes no explicit mention of a 0.8 threshold or any other threshold. As the DCWP states in a FAQ, ``the law does not require any specific actions based on the results of a bias audit, including ceasing the use of an AEDT shown to result in disparate impact" \citep{new_york_city_department_of_consumer_and_worker_protections_automated_2023}. The FAQ does note that other laws on employment discrimination still apply outside of the scope of LL 144, which would presumably include the EEOC’s definition of disparate impact \citep{new_york_city_department_of_consumer_and_worker_protections_automated_faq_2023,equal_employment_opportunity_commission_assessing_2023,code_of_federal_regulations_29_2024}.\footnote{The first case of algorithmic hiring bias—for age discrimination—was settled with the EEOC in 2023, but it was for a system that automatically screened out all older applicants and not a typical disparate impact scenario \citep{equal_employment_opportunity_commission_itutorgroup_2023, gilbert_eeoc_2023}.} 
Because LL 144 does not take a stance on permissible thresholds, it is also silent on remediation, and thereby limits the utility of the law for driving and measuring reduction of discrimination over time.  

Federal regulators have begun to set expectations about algorithmic systems and discrimination law that forbid employers from creating adverse impact through AI systems \citep{chopra_joint_2023, equal_employment_opportunity_commission_americans_2022, equal_employment_opportunity_commission_assessing_2023}, including holding hearings on the appropriateness of the four-fifths rule for AI use cases \citep{equal_employment_opportunity_commission_navigating_2023,webber_ai_2023}. Nonetheless, this guidance is situated in the EEOC’s long-standing decision tree for determining acceptable selection practices, which offers few hard lines amenable to auditing and relies heavily on judgment calls and precedents \citep{equal_employment_opportunity_commission_employment_2007}. Nor has any federal body established guidance about safe harbors for audits conducted in compliance with a local or state law. If introduced, such safe harbors would protect employers from disincentives to disclose information about discriminatory algorithms, as is the case in other areas of law \citep{unger_doj_2023}. Given the absence of any safe harbors, employers complying with LL 144 by posting audit reports might also be opening themselves to other types of liability. 

\subsection{Survey Response Rates} \label{surveyresponse}
Of all the employers in our sample, 29 listed no email or contact form or had a contact form with a character limit that was so low as to make the message incomplete. Of the emails we did send in the first round, 15 bounced or received a reply saying the address was not monitored.

We gave the employers 22 days to respond to our second outreach, from Dec 19 - Jan 10th. Among the 387 follow-up survey emails that we sent, 26 employers completed the survey, and 15 contacted us in some other way, either by email or phone. In total, this follow-up process yielded two audit reports that we had not previously observed.

\subsection{Student Investigators} \label{studentinvestigators}
We are deeply grateful to the students of COMM/INFO 2450 for their thorough and extensive collaboration on this project, among them Amelia Neumann, Andrew Wu, Angelina Chen, Anjiya Amlani, Anushka Shorewala, Bella Samtani, Bingsong Li, Carina Wang, Caroline Michailoff, Chelsea Lin, Chengling Zheng, Diana Flores Valdivia, Doan-Viet Nguyen, Dora Xu, Erik Starling, Evelyn C. Kim, Gianna Chan, Haley Qin, Hannah M. Yeh, Hermione Bossolina, Hope Best, Ingrid Gruener Luft, Jacob Levin, Jimin Kim, Jolene Ie, Kashmala Arif, Katherine Hahnenberg, Kathryn M. Papagianopoulos, Kevin Jianzhi Wang, Kexin Li, Kimmie Jimenez, Lili Mkrtchyan, Lindsay Peck, Maksym "Max" Bohun, Mark Timothy Bell, Mika Labadan, Minh H. Le, Neha Sunkara, Nicholas Bergersen, Nicholas Won, Nicole Tian, Noah Salzman, Nuo Cen, Omar Ahmed, Owen J. Chen, Reid Fleishman, Reinesse Wong, Sebastian Klein, Shukria Mirzaie, Simah Sahnosh, Siying Cui, Sophia Torres Lugo, Sritanay Vedartham, Subhadra Das, Thej Khanna, Varsha Gande, Weiyan Zhang, Wen Yu Chen, Yanran Li, Yiwen Zhang, Yubin Yang, Yuchen Yang, Yuyan Wu, and Zoey Arnold.

\end{document}